\documentclass[a4paper,11pt]{article}
\usepackage{pos}

\usepackage{amsmath}

\newcommand{\Regk}{\ensuremath{R_k}}

\title{Renormalization group approaches to Quantum Gravity and Tensions in Modern Cosmology}

\author*{Vasilios Zarikas}

\affiliation{Department of Mathematics, University of Thessaly,\\3rd Km Lamia-Athens, Lamia, Greece}

\emailAdd{vzarikas@uth.gr}

\FullConference{Corfu Summer Institute 2023 "School and Workshops on Elementary Particle Physics and Gravity" (CORFU2023)\\
 23 April - 6 May , and 27 August - 1 October, 2023\\
Corfu, Greece\\}

\abstract{

Asymptotic Safety is a promising framework towards the understanding, in a non-perturbative way, of Quantum Gravity. It treats the Newton’s constant $G_N$ and the cosmological constant $\Lambda$ as running coupling of an effective action. At the phenomenological point of view the values of $G_N$ and $\Lambda$ depend on the energy density of the system under consideration. This fact has interesting astrophysical and cosmological consequences. The present work discusses the effects on the distance measurements from SNIa light curves and on the strong and weak gravitational lensing measurements.
 }

\begin{document}
\maketitle

\section{Introduction}

Reported here are intriguing phenomenological implications stemming from $\Lambda$ varying gravity theories, inspired by certain quantum gravity models. The treatment presented in this study is broad-reaching, applying to various actions involving $\Lambda$ variation, particularly those employed in Renormalization Group (RG) approaches to quantum gravity. A pivotal aspect is an effective gravitational action featuring a Cosmological Constant (CC), $\Lambda$, that depends on the system's energy density thus its characteristics such as size, energy content etc. Notably, when the system encompasses astrophysical entities like galaxy clusters or black holes, significant corrections manifest in several observable parameters. These corrections appear in luminosity distance and strong/weak lensing measurements, among others.

Thus, the current analysis pertains to  models that the cosmological constant $\Lambda(x_i,t)$ can vary based on the physical attributes of the system in question, such as spatial dimensions and energy content i.e. energy density. Such variability is a characteristic feature of certain intriguing approaches to quantum gravity already established in the literature. When applied to astrophysical entities like galaxy clusters or black holes rather than the entire Universe, these models yield distinct phenomenological implications. 
We focus on obervational signatures concerning SNIa luminocity distances and weak and strong gravitational lensing.

\section{Asymptotic Safety}

Weinberg initially proposed in 1976 \cite{Weinberg:1980gg} that perturbative renormalizability isn't the sole solution for a theory's completeness at high energies. Instead, it suffices for the theory to possess finite values for a finite number of parameters at UV energies. Asymptotic Safety (AS) adheres to these principles, aiming for a theory characterized by a finite number of finite parameters with fixed values at high energies. Though the concept has been around for years, it wasn't until the late 1990s, following works by Wetterich and Reuter, that asymptotically safe framework was formally formulated, as detailed in \cite{Wetterich:1992yh, Reuter:1996cp}.

Asymptotic safety (AS), reviewed in \cite{Reuter:2019byg},\cite{Percacci:2017fkn},\cite{Platania:2022gtt},\cite{Bonanno:2020bil}, operates within the confines of four-dimensional spaces and continuous manifolds. It presents a minimalist approach, retaining the symmetries and fields of both Quantum Field Theory (such as the Standard Model) and General Relativity (GR). Moreover, it adopts a background-independent approach.

Notably, AS suggests that General Relativity and its extensions, including  some with higher derivatives in the Einstein-Hilbert action, can be models capable of non-perturbative renormalization. However, it should be noted that AS is not a complete framework due to the absence of knowledge about the Lagrangian governing our universe. 

AS theory defines field contents and symmetries. Subsequently, the family of actions is determined, specifying interactions of fields that abide by these symmetries. Functional renormalization group equations are then employed within the theory space, which encompasses all actions with "coordinates" coupling constants (e.g., $G$, $\Lambda$). The renormalization group, (RG), flow establishes connections between physics at different scales ($k$), resulting in running coupling constants such as $G$ and $\Lambda$.
The RG method eliminates degrees of freedom in the ultraviolet Lagrangian containing interaction terms multiplied by couplings. As the approach progresses towards lower energies, it describes the values of running couplings. Ultimately, it enables the determination of IR couplings at nearly zero energies, where the effective potential typically exhibits a complex structure.

The functional renormalization group (FRG) methodology is anchored on a scale-dependent effective action denoted as $\Gamma_k$. The parameter $k$ traverses from $\Gamma_{k\to\,\infty}$, where no quantum fluctuations have been integrated out, to $\Gamma_{k\to\,0}$, signifying the stage where all quantum fluctuations have been integrated away.
The FRG flow equation for $\Gamma_{k}$, reads \cite{Wetterich:1992yh}, \cite{Reuter:1996cp},

\begin{equation}\label{eq:floweq}
k\partial_k\,\Gamma_k=\frac{1}{2}\mathrm{sTr}\left[\left(k\partial_k\,\Regk\right)\left(\Gamma_k^{(2)}+\Regk\right)^{-1}\right]\,.
\end{equation}

In this context, the right-hand side integrates quantum fluctuations with momenta approximately of the scale $k$, which have the most significant impact on the variation of $\Gamma_k$ at that scale. The second-order term, $\Gamma_{k}^{(2)}$, represents a functional derivative with respect to all fields, alongside with the regulator functional $\Regk$ and the supertrace $\mathrm{sTr}$, which aggregates over all indices.

The truncation of the Effective Average Action to the Einstein-Hilbert form, initially utilized to demonstrate Asymptotic Safety, Eq(\ref{trunc}), leads to RG equations showcasing intriguing trajectories and a non-Gaussian fixed point, known as the Reuter fixed point. 
\begin{equation}
	\Gamma_k = \frac{1}{16\pi}\int d ^4x \sqrt{-g}\,G(k)^{-1}\Big(R-2\Lambda(k)\Big) + \text{gauge fixing and ghost terms}
	\label{trunc}
	\end{equation}
Although the RG scale dependence $k$ differs from the physical scale dependence, the structure of asymptotically safe models, through momentum-dependent correlation functions and form factors, gives rise to running couplings pertinent to the gravity sector, notably Newton's constant and the cosmological constant, $G(k)$, $\Lambda(k)$. Ultimately, observables are successfully defined based on these fundamental elements of the theory. A recent study also corroborated the physical graviton propagator's close resemblance to the propagator used in AS calculations, thereby justifying the physical running of $G$ and $\Lambda$ (\cite{Bonanno:2021squ}).

The RG flow for $G(k)$ and $\Lambda(k)$ in the Ultraviolet, UV, regime ($k\rightarrow\infty$) is given by ${G(k)}_{{UV}} = \frac{g_{*}}{k^{2}}$,  
$\Lambda\left( k \right)_{{UV}} = \ \lambda_{*}k^{2}$, 
where the dimensionless $(g_{*},\lambda_{*})$ take finite values (UV fixed point). 
This suggests a gravity anti-screening, potentially explaining phenomena near the Big Bang or during the concluding phases of gravitational collapse. These flow hold relevance solely within the trans-Planckian domain, with their following evolution governed by a system of partial differential equations. The dynamics of these coupling constants exhibit variability based on the 
composition of matter. 

Perhaps the most illuminating physical explanation in
this context has been presented by Polyakov,\cite{Polyakov:1993tp}, who noticed that as gravity is always attractive and therefore a larger cloud of virtual particles implies a stronger gravitational force, Newton's constant $G$ should be anti-screened at small distances. The implication of this behaviour suggests that the dimensionless coupling constant tends to a finite non-zero limit at small distances
When we go at higher energies is like looking closer a particle so we take under consideration smaller number of virtual particles. Thus, $G$ gets smaller. The same applies for a negative CC.
A positive cosmological constant term on the other hand, is always repulsive, therefore a larger cloud of virtual particles implies a less repulsive force, and the cosmological constant $\Lambda$ should be larger at small distances. The implication of this behaviour suggests that the dimensionless coupling constant $\lambda = \Lambda\,k^{-2}$ tends to a finite non-zero limit at small distances.

At lower energies, the dimensionless Newton constant and cosmological constant
 $(g, \lambda$) are running, \cite{Adeifeoba:2018ydh},
\begin{subequations}\label{ir1}
	\begin{align}
		&g(k)=g_*+g_1\left(\frac{k}{M_{d}}\right)^{-\theta_1}+g_2\left(\frac{k}{M_{d}}\right)^{-\theta_2}\,,\\
		&\lambda(k)=\lambda_*+\lambda_1\left(\frac{k}{M_{d}}\right)^{-\theta_1}+\lambda_2\left(\frac{k}{M_{d}}\right)^{-\theta_2}\,,
	\end{align}
\end{subequations}
where the critical exponents $\theta_i$ determine the flow behaviour.
A new energy scale is expected, named $M_d$. It is the energy scale where the ensemble of quantum spacetimes decohere and a classical spacetime is emmerged. $M_d$ is expected to be related with the Planck energy scale $M_P$. 
Equations (\ref{ir1}) hold within the infrared regime and may be pertinent on astrophysical scales. The parameters within these equations are constrained to meet experimental data and cosmological/astrophysical observations. To ensure a viable phenomenology in the late cosmological era, they must conform for example to the range elucidated in \cite{Zarikas:2017gfv,Anagnostopoulos:2018jdq,Anagnostopoulos:2022pxa, Zarikas:2023glu}.

Numerous efforts in high-energy physics, cosmology, and astrophysics seek to address various challenges utilizing the properties of Asymptotic Safe Gravity. These endeavors include attempts to explain inflation, dark energy, dark matter, and to reconcile discrepancies such as those relating to MOND dynamics and Hubble tensions, 
\cite{Zarikas:2017gfv}, 
\cite{Weinberg:2009wa},
\cite{Mitra:2021ahd}.
\cite{Bonanno:2020qfu}, 
\cite{Eichhorn:2022gku, Knorr:2022ilz}, \cite{Nielsen:2015una, Bonanno:2015fga, Gubitosi:2018gsl, Lehners:2019ibe,  Bentivegna:2003rr, Ahn:2011qt, Casadio:2010fw, Bonanno:2017zen}.

In AS phenomenology, $k$ denotes the reciprocal of a characteristic spatial length over which the fields of the system are averaged. Consequently, $k$ correlates with the scale of the physical system under examination. For astrophysical entities, $k$ may vary as a function of their proper length and, equivalently, their energy density \cite{Bonanno:2005mt,Bonanno:2017pkg}. 
An accurate treatment necessitates the non-perturbative computation of a finite-temperature effective action encompassing both the gravitational field and other fields, such as those in the Standard Model.

It is pertinent to highlight the relationship between $G$ and $\Lambda$ with the energy density within the AS framework. This dependency holds different implications in different areas of the Universe, such as voids, clusters or filaments as well as in the early Universe during the "Big Bang" phase \cite{Kofinas:2016lcz, Bonanno:2009nj, Bonanno:2010mk}, and in the cores of compact objects like stellar interiors, neutron stars, and black holes \cite{Torres:2014gta, Bonanno:2019ilz, Eichhorn:2022bgu, Kofinas:2015sna, Platania:2020lqb, Platania:2023srt}.

The aim of the paper is to propose ways to measure different values of the cosmological constant between regions of the Universe that are associated with different energy densities. 
In the third section we analyse the case of signals coming from distant supernovae SNIa, \cite{Linder:2019aip,Mitra:2022ykq}, while in the fourth section we describe the case of strong/weak lensing, \cite{AbhishekChowdhuri:2023ekr},\cite{Grespan:2023cpa}.

Our investigation maintains the assumption of a nearly constant value for Newton's constant $G$. This assumption aligns with phenomenologically viable scenarios of Renormalization Group flow towards the infrared limit, where $G$ exhibits only weak variation.

\section{Luminosity distances from signals traveling through voids and filaments/clusters}

The idea here is to measure different values of the cosmological constant utilizing SNIa signals traveling though areas of the Universe with as different as possible energy densities \cite{Good:2023xwp}. This means that we have to use two different samples. In the first sample we collect signals coming from areas with more filaments with clusters of galaxies while in the second sample we have to include signals passing mainly through voids.

The Hubble rate, $ H=\frac{-1}{(1+z)}\frac{dz}{dt}$ can be parameterized as
\begin{equation}\label{H}
H^2 = H_0^2 \left[ \Omega_M\,(1+z)^3 +\Omega_\gamma\,(1+z)^4 + \Omega_\Lambda + \Omega_k\,(1+z)^2 \right]   
\end{equation}
with the different $\Omega$s representing the matter, radiation, cosmological constant, and topological curvature contributions and $z$ the redshift..

The luminosity distance is $D_L=\sqrt{L/4\pi F}$ with $F$ the flux. 
The photon rate decreases by a factor of $a/a_0 = 1/(1+z)$, accompanied by an additional energy decrease by the same factor owing to gravitational redshift. Conventionally, we set $a(t=0)=1$.
Furthemore, since $F=a^2\frac{L}{4\pi\,S(r)^2}$ with the comoving distance $S(r) = \int_0^r \frac{dx}{\sqrt{1-k\,x^2}}$, the luminosity distance is given by

\begin{equation}\label{DL}
D_L = -(1+z)\int_{t_0}^{t_{emit}}\frac{dt}{a(t)}=-(1+z)\int_{a}^{1} \frac{da}{a^2\,H(a)} =(1+z)\int_{0}^{z} \frac{dz'}{H(z')} .
\end{equation}

In relation to the statistical analysis concerning distance measurements, we can explore various null hypotheses while working with luminosity distance as an observable.
We can compare the mean value of the observable, $D_L$, derived from observations with its theoretical counterpart assuming a preferred value for $\Lambda$ from theory. Estimating distances from Type Ia supernova magnitudes in the conventional manner, we can juxtapose these two $D_L$ values. Furthermore, another test involves constructing one sample with signals traversing significantly more voids along their line of sight compared to a second sample. Here, statistical differences in the mean $\Lambda$ value between the two samples would suggest potential quantum corrections to $\Lambda$.

We adopt for simplicity two distinct values for the Hubble rate. One value corresponds to the expansion rate observed in low-density systems like voids, denoted as $H_v$, while another value, denoted as $H_c$, characterizes the expansion rate in over-dense cosmic regions, such as filaments and clusters of galaxies.
In addition, we assume the same value, an average CC value, for all voids and a different, but same CC value among filaments/clusters to streamline the modeling process.

For a light beam (originated by a source at redshift $Z$) passing through $N$ voids and $M$ regions with filaments, the luminosity distance is approximatelly given by
\begin{eqnarray}\label{DLNΜ}
D_L = (1+z_{vf\,N}) \int_{0}^{Zvf\,N} H_{c}^{-1}\,dz' + \sum_{p}\,\frac{1+z_{vi\,p}}{1+z_{vf\,p}}\int_{Z_vf\,p}^{Zvi\,p} H_{v}^{-1}\,dz' + \nonumber \\
+ \sum_{q}\,\frac{1+z_{ci\,q}}{1+z_{cf\,q}}\int_{Z_cf\,q}^{Zci\,q} H_{c}^{-1}\, dz'+\frac{1+z}{1+z_{vi\,1}}\int_{Zvi\,1}^{Z} H_{c}^{-1}\,dz',
\end{eqnarray}
where $p=N,N-1,...1$ and $q=M-1,M-2,...2$ and with $z_{vi\,p}$, ($z_{ci\,q}$), the gravitational redshifts of entering the p-th void (q-th cluster), and $z_{vf\,p}$, ($z_{cf\,q}$) the redshift of light exiting this void (cluster) respectively.

If detailed knowledge regarding the distances between voids and clusters is inaccessible, an alternative approach is feasible. We can define two distinct samples: one predominantly comprising voids and the other emphasizing clusters or filaments. To implement this strategy, we gather a substantial dataset of signals traversing areas exhibiting over-densities and under-densities. Nonetheless, as the sample size increases, the discrepancy between densities diminishes, necessitating a trade-off.

Next, we analyze a situation where signals from a particular sample of astrophysical sources arrive on Earth after traveling through a region of space with lower-than-average matter density. Let's assume this region has an average matter density represented by $\rho_{un}$.
Next, let's examine a second set of signals that have passed through areas of higher-than-average density, indicated by an average matter density of $\rho_{ov}$. As a result, we can investigate statistical differences in the values of $\Lambda$.
Then, luminosity distance for signals coming through under-density regions is given by
\begin{equation}\label{DLrho}
D_{un}  = (1+z)\int_{0}^{z} \frac{dz'}{H_{un}(z')} ,
\end{equation}
with  
\begin{equation}
H_{un}^2 = H_0^2 \left[ \frac{\rho_{un}}{\rho_{cr}}\,(1+z)^3  + \frac{\Lambda_{un}}{3H_0^2} + \Omega_k\,(1+z)^2 \right],  
\end{equation}
Similarly we can evaluate the luminocity distance through over-density regions from the formula $D_{ov}= (1+z)\int_{0}^{z} \frac{dz'}{H_{ov}(z')} $ related to $\Lambda_{ov}$ from
\begin{equation}
H_{ov}^2 = H_0^2 \left[ \frac{\rho_{ov}}{\rho_{cr}}\,(1+z)^3  + \frac{\Lambda_{ov}}{3H_0^2} + \Omega_k\,(1+z)^2 \right]. 
\end{equation}
Now according to the AS theory, \cite{Bonanno:2005mt},\cite{Bonanno:2017pkg}, we can approximately write
\begin{equation}
\Lambda_{un} \propto \xi\,\rho_{un}^{1/2}.       
\end{equation}
Certainly, a strong test for RG approaches to quantum gravity would be to verify the ratio between the two values of $\Lambda$ in addition to the statistical difference of the mean/median values of the $\Lambda$s.
\begin{equation}
\frac{\Lambda_{un}}{\Lambda_{ov}}=\left(\frac{\rho_{un}}{\rho_{ov}}\right)^{1/2}.
\end{equation}
A non homogeneous treatment of the whole Universe would be more appropriate and will be presented in a forthcoming paper. This work aims to describe mainly the phenomenon.

\section{Implications for Strong and Weak Lensing measurements}

Approaches to quantum gravity that posit a discrete spacetime or an initial network of events can naturally address the issue of spacetime singularities. AS represents a theory that encompasses a continuous gravity field. The capability of AS to avoid singularities arises from the anti-screening of gravity strength and the potential presence of a positive $\Lambda$ in the UV regime. Additionally, Asymptotic Safe Gravity alters the characteristics of black holes, such as a non-singular center, an inner horizon, potential final remnants, and radiation, as proposed in 
\cite{Eichhorn:2022bgu, Platania:2023uda, Falls:2010he, Cai:2010zh, Saueressig:2015xua}.

AS corrects at tree level Schwarzschild-de Sitter metric with a homogeneous isotropic metric. Both metrics contain $k$ dependent cosmological and Newton constants.	
AS inspired quantum improved Schwarzschild-de Sitter metric Eq.(\ref{ASBH}), \cite{Koch:2014cqa}. These type of metrics can be chosen to describe galaxies and/or clusters of galaxies.
We have
\begin{equation}\label{ASBH}
ds^2=-\Big(1-\frac{2G_k M}{R}-\frac{1}{3}\Lambda_k R^2\Big)dT^2
+\frac{dR^2}{1-\frac{2G_k M}{R}-\frac{1}{3}\Lambda_k R^2}
+R^2\,d\Omega^2\,,
\end{equation}
where $G_k$ and $\Lambda_k$ are functions of a
characteristic scale $k$ of the system under investigation.

AS theory at the present incomplete stage cannot provide the exact dependence of the cutoff $k$ on physical scales like length or energy. For phenomenological purpose various simple scalings have been proposed 
\cite{Bonanno:2000ep}, \cite{Bonanno:2019ilz},
\cite{Bonanno:2007wg},  \cite{Bonanno:2017pkg}. For astrophysical black holes $k$ is associated with the characteristic astrophysical length scale $L$, the radius, $R$. So, the ansatz for $k$ is $k=\xi/R$, with $\xi$ is a dimensionless order-one number. However, a better natural choice which generates the desired phenomenology, is to set as $L$ equal to the proper distance $D>0$.  The use of proper distances proved also to be a sucesful choice also for excibiting a singularity avoidance/smoothing in spherical solutions of AS gravity
\cite{Kofinas:2015sna}.

The proper distance along a radial path, $dT=d\theta=d\varphi=0$ from $R_0$ till $R$ is given by
\begin{equation}
D(R)=\int_{R_{0}}^{R}\frac{d\mathcal{R}}{\sqrt{F}}\,.
\label{proper}
\end{equation}
with $F=1-\frac{2G_k M}{R_S}-\frac{1}{3}\Lambda_k R_S^2$.
In the IR regime it is expected that the dimensionless Newton constant and cosmological constant run according to the trajectory
\begin{equation} \label{RGIR}
 g_{IR}\left( k \right) = g_{*} + h_{1}k^{\theta_{1}}, \quad \quad
	\lambda_{IR}\left( k \right) = \ \lambda_{*} + h_{2}k^{\theta_{2}}, 
\end{equation}
with $(\theta_{1},\theta_{2} \geq 0)$ two unknown critical exponents. 

Now, the gravitational lensing from a point mass is estimated using the gravitational potential (energy per unit mass) which is 
\begin{equation} \label{pot}
    \Phi(R,x)=-\frac{G_k\,M}{\sqrt{R^2+x^2}}-\frac{1}{6}\Lambda_k (R^2+x^2)
\end{equation}
It should be noted that the second term in Eq.(\ref{pot}), is repulsive if AS generates a positive CC. However, there are some AS studies suggesting that the inclusion of Standard Model (SM) gauge fields (and fields beyond the SM) can cause to a negative CC value or a value for CC that changes sign during the RG flow. 
The deflection angle relevant for point mass lensing is approximately equal to two times the newtonian prediction, 
\begin{equation}
  \hat{a} =2 \int _{-\infty}^{\infty} \frac{\partial\, \Phi(R, x)}{\partial\, R}  dx  
\end{equation}
Here, $G_k$ and $\Lambda_k$ are expected to depend on the distance from the center of the spherical object. Following AS theory suggestions we can approximately write that 
\begin{equation}
 k \propto \frac{1}{D(R)}   
\end{equation}
The dependence of $G_k$ and $\Lambda_k$ on $k$ is determined by the RG partial differential equations and approximately can be found using Eqs.(\ref{RGIR}).
Since the deflection angle is non trivially modified, it is obvious that both strong and weak lensing measurements should be processed under this new perspective and the corresponding inferences will be different. 

Images of quasars, when lensed, appear as point-like due to their vast distances from Earth, while galaxy images are extended. The latter case can lead to the blending of individual lensed images, resulting in visually captivating structures that can be challenging to definitively identify as products of lensing. Strong lensing provides insights into the matter distribution within galaxies and clusters of galaxies, causing background sources to be split into multiple images. Since lensing is influenced by both dark and luminous matter, whereas other methods primarily focus on the luminous component, comparing results from different techniques allows us to deduce the distribution of the elusive dark matter. Additionally, examining lenses within their cosmological framework can help constrain cosmological parameters.

On another note, lensing can also modify the shapes of background galaxies without generating multiple images. This phenomenon, termed weak lensing, cannot be determined from a single observation but becomes evident when multiple sources are studied, producing a statistically notable effect. Even if objects behind the lens are positioned far from its central point, their light can still be deflected by a sufficiently massive lens. If the lens's surface mass density exceeds a specific threshold, multiple images are produced; otherwise, a subtle distortion of the background source occurs.
Interestingly, weak lensing not only allows us to investigate individual mass concentrations but also offers insights into the universe's large-scale structure between the observer and the source. This aspect is influenced by cosmological factors like the Hubble constant, the universe's total energy density, and the relative proportions of radiation, matter, and dark energy. 

In a forthcoming publication, lensing datasets will be analysed and presented assuming asymptotic safe gravity which for the case of lensing requires AS corrections to be taken into account both for black hole solutions and for the cosmological background.

\section{Conclusions}

In this study, I propose two specific observational signatures to test asymptotic safe gravity models inspired by RG approaches to quantum gravity.

If the cosmological constant, $\Lambda$, varies due to quantum corrections and depends on the energy density of the astrophysical system being studied, it leads to both astrophysical and cosmological implications. The luminosity distance is considered as an observable, and a technique has been suggested to investigate the idea that $\Lambda$ may have distinct values in cosmic voids compared to those in filaments and clusters.

Furthermore, the case of gravitational lensing was sketched. The deflection angle for a point source has been estimated and it was shown that is modified in the context of Asymptotic Safety. The fact that $G$ varies, alters the gravitational potential that bends the light. Moreover, and perhaps more significantly, there is an additional repulsive or attractive force due the existence of a new potential term arising from the varying $\Lambda$.

\newpage
\bibliography{ref.bib}

\end{document}